\documentclass[showpacs,prb,
twocolumn,
]{revtex4}
\usepackage{graphicx}
\usepackage[dvips]{color}
\begin{document}
\title{%
Crossover from bias-induced to field-induced breakdowns in
one-dimensional band and Mott insulators attached to electrodes}
\date{\today}

\author{Yasuhiro Tanaka}
\email{yasuhiro@ims.ac.jp}
\author{Kenji Yonemitsu}
\affiliation{Institute for Molecular Science, Okazaki 444-8585, Japan}
\affiliation{Department of Functional Molecular Science, Graduate
University for Advanced Studies, Okazaki 444-8585, Japan}
\affiliation{JST, CREST, Sanbancho, Chiyoda-ku, Tokyo 102-0075, Japan}

\begin{abstract}
Nonequilibrium states induced by an applied bias voltage $(V)$ and the
 corresponding current-voltage characteristics of one-dimensional
 models describing band and Mott insulators are investigated
 theoretically by using nonequilibrium Green's functions. We attach the
 models to metallic electrodes whose effects are incorporated into the
 self-energy. Modulation of the electron density and the scalar
 potential coming from the additional long-range interaction are
 calculated self-consistently within the Hartree approximation. For both
 models of band and Mott insulators with length $L_C$, the bias voltage
 induces a breakdown of the insulating state, whose threshold shows a
 crossover depending on  $L_C$. It is determined basically by the bias 
$V_{\rm th}\sim \Delta$ for $L_C$ smaller than the correlation length
 $\xi=W/\Delta$ where $W$ denotes the bandwidth and $\Delta$ the energy
 gap. For systems with $L_C\gg \xi$, the threshold is governed by the
 electric field, $V_{\rm th}/L_C$, which is consistent with a
 Landau-Zener-type breakdown, $V_{\rm th}/L_C\propto \Delta^2/W$. 
We demonstrate that the spatial dependence of the scalar potential is
 crucially important for this crossover by showing the case without
 the scalar potential, where the breakdown occurs at 
$V_{\rm th}\sim \Delta$ regardless of the length $L_C$.
\end{abstract}

\maketitle

\section{Introduction}
Nonlinear conduction in correlated electron systems such as
one-dimensional Mott insulators\cite{Taguchi_PRB00,Kumai_SCI99} and
two-dimensional charge-ordered
materials\cite{Yamanouchi_PRL99,Sawano_NATURE05,Kondo_JPSJ07,
Niizeki_JPSJ08,Sawano_JPSJ09,Inada_PRB09}
has been of great interest in the past few decades. They offer
intriguing subjects of nonequilibrium physics in condensed matter and
possibility for novel functions of electronic devices. For example, in a
typical quasi-one-dimensional Mott insulator,
Sr$_2$CuO$_3$,\cite{Taguchi_PRB00} a dielectric breakdown has been
observed experimentally by applying a strong electric field. A
dielectric breakdown has been reported also in an organic spin-Peierls
insulator,
K-TCNQ\cite{Kumai_SCI99} [TCNQ=tetracyanoquinodimethane]. For another
organic compound, (BEDT-TTF)(F$_2$TCNQ)
[BEDT-TTF=bis(ethylenedithio)tetrathiafulvalene], which is a
quasi-one-dimensional Mott insulator, metal-insulator-semiconductor
field-effect transistor device structures have been
reported,\cite{Hasegawa_PRB04} where its field-effect characteristics
are different from those of band
insulators.\cite{Yonemitsu_JPSJ05,Yonemitsu_PRB07,Yonemitsu_JPSJ09}

So far, theoretical investigations on nonlinear conduction for 
interacting electron systems that are initially insulating in their
equilibrium states have been done by several
authors.\cite{Yonemitsu_JPSJ05,Yonemitsu_PRB07,Yonemitsu_JPSJ09,
Oka_PRL03,Oka_PRL05,Oka_PRL05b,Oka_PRB10,Onoda_PTP06,Sugimoto_PTP07,
Sugimoto_PRB08,Okamoto_PRB07,Okamoto_PRL08,Ajisaka_PTP09}
In general, these studies are classified into two approaches depending
on whether an external force is written as an electric
field\cite{Oka_PRL03,Oka_PRL05,Oka_PRB10,Onoda_PTP06,Sugimoto_PTP07,Sugimoto_PRB08} 
or a bias
voltage.\cite{Yonemitsu_JPSJ05,Oka_PRL05b,Yonemitsu_PRB07,Yonemitsu_JPSJ09,
Okamoto_PRB07,Okamoto_PRL08,Ajisaka_PTP09} The former approach is to
consider electron systems without
electrodes. The electric field is usually applied with open boundary
condition,\cite{Oka_PRL05} or equivalently with periodic boundary
condition by using a time-dependent magnetic
flux.\cite{Oka_PRL03,Oka_PRB10} In one dimension, Oka and Aoki studied
the Hubbard model under a strong electric field by the time-dependent
density-matrix-renormalization-group method.\cite{Oka_PRL05} One of
their important results is that the dielectric breakdown of Mott
insulators is interpreted as a many-body counterpart of the Landau-Zener
(LZ) breakdown.\cite{Landau_SOW32,Zener_PRSL32}
This is known to describe the breakdown of band insulators
where the one-particle picture holds. The threshold is given as 
$E_{\rm th}\propto \Delta^2$ with $\Delta$ being an energy gap.
A mean-field approach to
electric-field-induced insulator-to-metal transitions by using Keldysh
Green's functions has been reported in
Ref. \onlinecite{Sugimoto_PRB08}. 

The latter approach is to consider an insulator attached to
electrodes. Interface structures must be explicitly taken into account
in real electric devices. 
Along this line, Okamoto investigated the current-voltage ($I$-$V$)
characteristics of heterostructures that consist of Mott-insulator
layers sandwiched by metallic leads by combining the dynamical
mean-field theory with the Keldysh Green's function
technique.\cite{Okamoto_PRB07,Okamoto_PRL08} Ajisaka {\it et al.}
studied the $I$-$V$ characteristics of an electron-phonon system coupled
with two reservoirs by a field-theoretical method.\cite{Ajisaka_PTP09}
In the studies listed above, Okamoto discussed the $I$-$V$
characteristics in the framework of the LZ breakdown, while Ajisaka 
{\it et al.} proposed a different mechanism with the threshold bias
voltage $V_{\rm th}\sim \Delta$. These results seem to be inconsistent
with each other. 

When we consider a nanostructure in which some material is attached to
left and right metallic electrodes, the bias voltage $V$ applied to the
material is described as $V=\mu_L-\mu_R$ with $\mu_L$ and $\mu_R$ being
the chemical potentials
of the left and right electrodes, respectively. Here we assume that
the work-function difference at the interfaces is absent for
simplicity. In this case, one might expect that the current flows
when some energy levels of the material appear in the region between
$\mu_L$ and $\mu_R$, indicating the threshold governed by the applied
bias voltage $V_{\rm th}\sim \Delta$. In fact, for the transport in a
field-effect transistor with a small
channel, such an explanation has been used frequently.\cite{Datta_05}
However, for a bulk insulator with an energy gap $\Delta$, this picture
does not hold and should be replaced by the LZ mechanism where the
threshold is determined by the electric field, 
$E_{\rm th}\propto \Delta^2$. This consideration poses us a question
about which parameter determines
the mechanism of the breakdown. In particular, we address the
condition for the realization of the LZ breakdown in a structure
with electrodes. It is also important to know the relation between the
approaches using the structure with electrodes and those which do not
include them explicitly. We point out that the size of an insulator as
well as the potential distribution inside it determine the nature of
the breakdown. 

In this paper, we study one-dimensional band and Mott insulators
attached to two electrodes (see Fig. \ref{fig:model}), using the
nonequilibrium Green's function
approach that has previously been used to discuss the suppression of
rectification at metal-Mott-insulator interfaces.\cite{Yonemitsu_JPSJ09}
This approach more naturally describes nonequilibrium steady states than
the approach based on the time-dependent Schr$\ddot{\rm o}$dinger
equation because a current oscillation is inevitable in the latter owing
to finite-size effects. The present method can be easily applied to
higher-dimensional systems. Preliminary results for the $I$-$V$
characteristics of two-dimensional charge-ordered systems are reported
in Ref. \onlinecite{Tanaka_PHYSB10}. 

We show that the applied bias voltage $V$ induces a breakdown of band
and Mott insulators at zero temperature. For both insulators, the
threshold shows a crossover as a function of the size of the insulating
region $L_C$. For systems with $L_C$ smaller than the correlation
length $\xi$, i.e., the characteristic decay length of the wave function
in the insulator, the breakdown takes place when the bias $V$ exceeds the
energy gap $\Delta$. For $L_C\gg \xi$, it is governed by the electric
field, $V/L_C$, which is consistent with the LZ tunneling
mechanism. Whether the charge gap is produced by the band structure or
by the electron-electron interaction is irrelevant to the crossover
phenomenon. We will focus on the spatial modulation of the wave
functions. 
For the crossover behavior and the deformation of the wave functions, 
the spatial dependence of the scalar potential inside the band or Mott
insulator is important. This is demonstrated in the Appendix by showing
the case without the scalar potential, where the bias-induced transition
occurs at $V_{\rm th}\sim\Delta$ regardless of the length $L_C$.

\section{Model and Formulation}
\begin{figure}
\includegraphics[height=1.1cm]{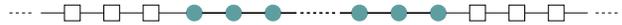}
\caption{(Color online) Schematic view of the model. A one-dimensional
 band or Mott insulator in the central part is connected by left and right
 electrodes. Solid (open) symbols represent the sites in the central
 part (electrodes).}
\label{fig:model}
\end{figure}

We consider a one-dimensional insulator that is attached to
semi-infinite metallic electrodes on the left and right sides as shown
in Fig. \ref{fig:model}. The insulator is referred to as the central
part, and described by the Hubbard model for a Mott insulator and a
tight-binding model with alternating transfer integrals for a band
insulator, both at half filling: 
\begin{eqnarray}
H&=&-\sum_{i=1}^{L_C-1}\sum_{\sigma}[t+(-1)^{i-1}\delta t)]
(c^{\dagger}_{i\sigma}c_{i+1\sigma}+h.c.)\nonumber \\
&+&U\sum_{i}(n_{i\uparrow}-\frac{1}{2})
 (n_{i\downarrow}-\frac{1}{2})\nonumber \\
&+&\sum_{\langle\langle
 ij\rangle\rangle}V_{ij}(n_i-1)(n_j-1),
\label{eq:ham}
\end{eqnarray}
where $c^{\dagger}_{i\sigma} (c_{i\sigma})$ denotes the creation
(annihilation) operator for an electron with spin $\sigma$ at the $i$th
site, $n_{i\sigma}=c^{\dagger}_{i\sigma}c_{i\sigma}$, and 
$n_i=n_{i\uparrow}+n_{i\downarrow}$. $L_C$ is the total number of sites
in the central part. The parameter
$t$ denotes the transfer integral, $\delta t$ its modulation, and $U$
the on-site interaction. 
We use $t$ as a unit of energy. For $\delta t=0$ and $U>0$, the first
and second terms in Eq. (\ref{eq:ham}) become the one-dimensional
repulsive Hubbard model, whereas they describe a band insulator for
$\delta t\neq 0$ and $U=0$. 
 The long-range Coulomb interaction term with
$V_{ij}=V_p/|i-j|$ is introduced because it is responsible for the
potential modulation near the metal-insulator interfaces ($i\gtrsim 1$
and $i\lesssim L_C$). The $V_{ij}$ term is treated by the Hartree
approximation, which is equivalent to the introduction of a scalar
potential that satisfies the Poisson equation. 
Here $\langle\langle ij\rangle\rangle$ means the summation over pairs of
the $i$th and $j$th sites with $i\neq j$ in the central part 
($1\leq i,j\leq L_C$). We briefly review our
formulation\cite{Yonemitsu_JPSJ09} below to treat steady states with a
finite voltage. 

For the metallic electrodes, we consider noninteracting electrons. 
The effects of the left and right $(\alpha=L,R)$ electrodes on the
central part are then described by the retarded
self-energies.\cite{Meir_PRL92,Wingreen_PRB93,Jauho_PRB94} 
For simplicity, we take the wide-band limit so that the self-energies
are independent of energy. In the present case, they become diagonal
matrices\cite{Jauho_PRB94}
\begin{equation}
(\Sigma^{r}_{\alpha})_{ij}=-\frac{i}{2}(\Gamma_{\alpha})_{ij}=-\frac{i}{2}\gamma_{\alpha}
\delta_{ii_{\alpha}}\delta_{ji_{\alpha}},
\end{equation}
where $\delta_{ii_{\alpha}}$ and $\delta_{ji_{\alpha}}$ are the
Kronecker delta and $i_L=1$ ($i_R=L_C$) denotes the site connected with
the left (right) electrode. We consider the case of
$\gamma_L=\gamma_R$. Within the Hartree-Fock approximation for the
on-site term in
Eq. (\ref{eq:ham}), the retarded Green's function for spin $\sigma$ is
written as
\begin{equation}
[G^{r}_{\sigma}(\epsilon)^{-1}]_{ij}=\epsilon \delta_{ij}-(H^{r}_{\rm
 HF\sigma})_{ij},
\end{equation}
with
\begin{equation}
(H^{r}_{\rm HF\sigma})_{ij}=(H_{\rm
 HF\sigma})_{ij}+\sum_{\alpha=L,R}(\Sigma^{r}_{\alpha})_{ij}.
\end{equation}
The off-diagonal elements of $(H_{\rm HF\sigma})_{ij}$ come from the
first term of Eq. (\ref{eq:ham}) and the diagonal elements are written
as
\begin{equation}
(H_{\rm HF\sigma})_{ii}=\psi_i+U\langle n_{i\bar{\sigma}}-1/2\rangle ,
\end{equation}
with $\bar{\sigma}=-\sigma$. Here, the scalar potential $\psi_i$ is
defined by the Hartree approximation to the long-range Coulomb
interaction as
\begin{equation}
\psi_i=\sum_{j\neq i}V_{ij}(\langle n_j\rangle -1)+ai+b, 
\label{eq:psi}
\end{equation}
with $a$ and $b$ being constants. These constants are so determined that
$\psi$ satisfies the boundary conditions:
\begin{eqnarray}
\psi_i=\left\{
\begin{array}{l}
\frac{V}{2}\ {\rm for}\ i=1\\
-\frac{V}{2}\ {\rm for}\ i=L_C,
\end{array}
\right.
\end{eqnarray}
for the bias voltage ${\it V}$. When $V$ is
positive, the left electrode has a higher potential for the electrons
and the current flows from left to right.\cite{Yonemitsu_JPSJ09} Here we
assume that the work-function difference is absent at the interfaces. 
By diagonalizing the complex symmetric matrix $H^{r}_{\rm HF\sigma}$,
the retarded Green's function is obtained as
\begin{equation}
[G^{r}_{\sigma}(\epsilon)]_{ij}=\sum_{m}\frac{u^{\sigma}_{m}(i)u^{\sigma}_{m}(j)}{\epsilon
 -E^{\sigma}_m}, 
\label{eq:gr}
\end{equation}
where $E^{\sigma}_m \equiv \epsilon^{\sigma}_{m}-i\gamma^{\sigma}_{m}/2$ 
($\epsilon^{\sigma}_m$ and $\gamma^{\sigma}_m$ are real) is the eigenvalue
of $H^{r}_{\rm HF\sigma}$ and $u^{\sigma}_{m}(i)$ is the corresponding
right eigenvector.

The electron density is calculated by decomposing it into the
``equilibrium'' and ``nonequilibrium'' parts\cite{Yonemitsu_JPSJ09} as
\begin{equation}
\langle n_{i\sigma}\rangle = n^{\rm eq}_{i\sigma}+\sum_{\alpha}\delta
 n^{\alpha}_{i\sigma}.
\label{eq:n_parts}
\end{equation}
The ``equilibrium'' part is defined by integrating the local density of
states as
\begin{equation}
n^{\rm eq}_{i\sigma}\equiv
 -\frac{1}{\pi}\int^{\infty}_{-\infty}d\epsilon
 [G^{r}_{\sigma}(\epsilon)]_{ii} f_C(\epsilon),
\label{eq:neq}
\end{equation}
where $f_C(\epsilon)=\theta (\mu_C-\epsilon)$ is the Fermi distribution
function with the chemical potential $\mu_C$ at the midpoint of the
right and left chemical potentials,
$\mu_C=(\mu_R+\mu_L)/2$. Since $\mu_L=V/2$ and $\mu_R=-V/2$, we have
$\mu_C=0$. For $V>0$, where the left chemical potential is higher than
the right, $\delta n^{L}_{i\sigma}$ ($\delta n^{R}_{i\sigma}$) is
interpreted as the inflow (outflow).

The ``nonequilibrium'' part of the
density $\delta n^{\alpha}_{i\sigma}$ is obtained from the
``nonequilibrium'' part of the lesser Green's function:
\begin{equation}
\delta n^{\alpha}_{i\sigma}\equiv \int^{\infty}_{-\infty}d\epsilon
 [\delta G^{<\alpha}_{\sigma}(\epsilon)]_{ii}.
\label{eq:nneq}
\end{equation}
In order to obtain $\delta G^{<\alpha}_{\sigma}(\epsilon)$, we first
decompose the lesser self-energy in the wide-band
limit\cite{Jauho_PRB94} as in Eq. (\ref{eq:n_parts}):
\begin{eqnarray}
\Sigma^{<}_{\sigma}(\epsilon)&=&i(\Gamma_L f_L(\epsilon)+\Gamma_R
 f_R(\epsilon))\nonumber \\
&=&\Sigma^{<{\rm eq}}_{\sigma}(\epsilon)+\sum_{\alpha}\delta
\Sigma^{<\alpha}_{\sigma}(\epsilon),
\end{eqnarray}
with
\begin{equation}
\Sigma^{<{\rm eq}}_{\sigma}(\epsilon)=i(\Gamma_L +\Gamma_R)f_C(\epsilon),
\end{equation}
and
\begin{equation}
\delta \Sigma^{<\alpha}_{\sigma}(\epsilon) =
 i\Gamma_{\alpha}[f_{\alpha}(\epsilon)-f_C(\epsilon)],
\end{equation}
where $f_{\alpha}(\epsilon)=\theta (\mu_{\alpha}-\epsilon)$. Then, we employ the
Keldysh equation
\begin{equation}
\delta G^{<\alpha}_{\sigma}(\epsilon)=G^{r}_{\sigma}(\epsilon)\delta
 \Sigma^{<\alpha}_{\sigma}(\epsilon)G^{a}_{\sigma}(\epsilon),
\end{equation}
where $G^{a}_{\sigma}(\epsilon)$ is the Hermitian conjugate of
$G^{r}_{\sigma}(\epsilon)$.
The expressions for 
$n^{\rm eq}_{i\sigma}$ and $\delta n^{\alpha}_{i\sigma}$, with which the
numerical calculations are carried out, are obtained by substituting
Eq. (\ref{eq:gr}) into Eqs. (\ref{eq:neq}) and
(\ref{eq:nneq}). The results are
\begin{equation} 
n^{\rm eq}_{i\sigma}=\sum_m {\rm
 Re}[u^{\sigma}_{m}(i)]^2\Bigl[\frac{1}{\pi}\tan^{-1}\frac{2(\mu_C-\epsilon^{\sigma}_m)}
{\gamma^{\sigma}_{m}}+\frac{1}{2}\Bigr],
\label{eq:neq2}
\end{equation}
and 
\begin{eqnarray}
\delta n^{\alpha}_{i\sigma}&=&\frac{\gamma_{\alpha}}{2\pi}
\int^{\infty}_{-\infty}d\epsilon
|[G^{r}_{\sigma}(\epsilon)]_{ii_{\alpha}}|^2
[f_{\alpha}(\epsilon)-f_C(\epsilon)]\nonumber \\
&=&
\frac{\gamma_{\alpha}}{2\pi}\sum_{n,m}\Bigl\{
{\rm Im}
\Bigl[
\frac{u^{\sigma}_m(i)u^{\sigma}_m(i_{\alpha})u^{\sigma *}_n(i)u^{\sigma
*}_n(i_{\alpha})}
{\epsilon^{\sigma}_m-\epsilon^{\sigma}_n-i\gamma^{\sigma}_m/2-i\gamma^{\sigma}_n/2}
\Bigr]
\Bigr\}\nonumber \\
&\times& \Bigl[
\tan^{-1}\frac{2(\mu_{\alpha}-\epsilon^{\sigma}_m)}{\gamma^{\sigma}_m}
-\tan^{-1}\frac{2(\mu_{C}-\epsilon^{\sigma}_m)}{\gamma^{\sigma}_m}\nonumber \\
&+&\tan^{-1}\frac{2(\mu_{\alpha}-\epsilon^{\sigma}_n)}{\gamma^{\sigma}_n}
-\tan^{-1}\frac{2(\mu_{C}-\epsilon^{\sigma}_n)}{\gamma^{\sigma}_n}
\Bigr]\nonumber \\
&+&{\rm Re}
\Bigl[
\frac{u^{\sigma}_m(i)u^{\sigma}_m(i_{\alpha})u^{\sigma *}_n(i)u^{\sigma
*}_n(i_{\alpha})}
{\epsilon^{\sigma}_m-\epsilon^{\sigma}_n-i\gamma^{\sigma}_m/2-i\gamma^{\sigma}_n/2}
\Bigr]\nonumber \\
&\times &
\Bigl[
\frac{1}{2}
\ln
\frac{(\mu_{\alpha}-\epsilon^{\sigma}_m)^2+(\gamma^{\sigma}_m/2)^2}
{(\mu_C-\epsilon^{\sigma}_m)^2+(\gamma^{\sigma}_m/2)^2}\nonumber \\
&-&
\frac{1}{2}
\ln
\frac{(\mu_{\alpha}-\epsilon^{\sigma}_n)^2+(\gamma^{\sigma}_n/2)^2}
{(\mu_C-\epsilon^{\sigma}_n)^2+(\gamma^{\sigma}_n/2)^2}
\Bigr].
\label{eq:nneq2}
\end{eqnarray}
In the above equations, we recover the electron density in the
equilibrium state without the electrodes if $\gamma^{\sigma}_m=0$,
since the bracket in Eq. (\ref{eq:neq2}) is reduced to the step function
and $\delta n^{\alpha}_{i\sigma}=0$.

The current from the left electrode is expressed by the
``nonequilibrium'' part of the density as\cite{Haug,Yonemitsu_JPSJ09} 
\begin{eqnarray}
J&=&\int^{\infty}_{-\infty}\frac{d\epsilon}{2\pi}\sum_{\sigma}{\rm Tr}
[\Gamma_L G^{r}_{\sigma}(\epsilon)\Gamma_R G^{a}_{\sigma}(\epsilon)]
[f_L(\epsilon)-f_R(\epsilon)]\nonumber \\
&=&\frac{\gamma_L\gamma_R}{2\pi}\int^{\infty}_{-\infty}d\epsilon
\sum_{\sigma}|[G^{r}_{\sigma}(\epsilon)]_{i_Li_R}|^2[f_L(\epsilon)-f_R(\epsilon)]\nonumber
\\
&=&\gamma_{R}\sum_{\sigma}\delta
 n^{L}_{i_R\sigma}-\gamma_{L}\sum_{\sigma}\delta n^{R}_{i_L\sigma},
\label{eq:cur}
\end{eqnarray}
where we set $e=\hbar=1$. 

\section{Results}
In this section, we show the results of $I$-$V$ characteristics, charge
densities, and the spatial dependence of wave functions for band
and Mott insulators. For both models, a breakdown of the insulating
state takes place when the bias $V$ becomes sufficiently large. The
threshold shows a crossover behavior as a function of the size of
the central part $L_C$, which indicates the mechanism of the breakdown
changes according to $L_C$. The profile of the scalar potential $\psi_i$
has crucial importance on the way of the breakdown. This is demonstrated
in the Appendix by showing that, if $\psi_i$ is artificially set to zero
for all $i$, the crossover phenomenon disappears.

\subsection{Band insulators}
\begin{figure}
\includegraphics[height=5.7cm]{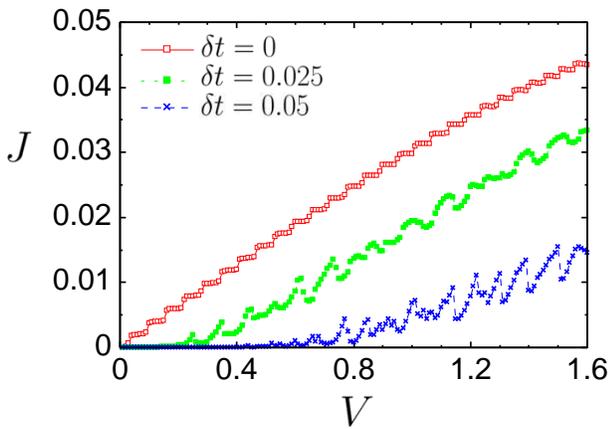}
\caption{(Color online) $I$-$V$ characteristics of one-dimensional band
 insulators for $\delta t=0.025$ and $\delta t=0.05$. The other
 parameters are $L_C=200$, $U=0$, $V_p=0.1$, and $\gamma_L =\gamma_R
 =0.1$. Results for $\delta t=0$ are also shown where we set $L_C=200$,
 $U=0$, $V_p=0$, and $\gamma_L =\gamma_R =0.1$.}
\label{fig:band_iv1}
\end{figure}

Figure \ref{fig:band_iv1} shows the $I$-$V$ characteristics for band
insulators with
$\delta t=0.025$ and $0.05$. The other parameters are $L_C=200$, $U=0$,
$V_p=0.1$, and 
$\gamma_L =\gamma_R = 0.1$. For comparison, we show the results for the
regular transfer integrals ($\delta t=0$) with $L_C=200$, $U=0$,
$V_p=0$, and 
$\gamma_L =\gamma_R = 0.1$. For $\delta t=0$, the current $J$ becomes
nonzero for $V\neq 0$ since the central part is metallic. The $I$-$V$
curve has stepwise structures owing to the finite-size effect. 
For $V=0$ and $\delta t>0$, the central part is a band insulator with
the energy gap $\Delta=4\delta t$. Because of the gap, $J$ is suppressed
near $V=0$. The $I$-$V$ curves for finite $\delta t$ show more complex
structures than that for $\delta t=0$. Apart from the fine
structures, $J$ increases almost linearly for large $V$, which indicates
a breakdown of the band insulator.

\begin{figure}
\includegraphics[height=5.3cm]{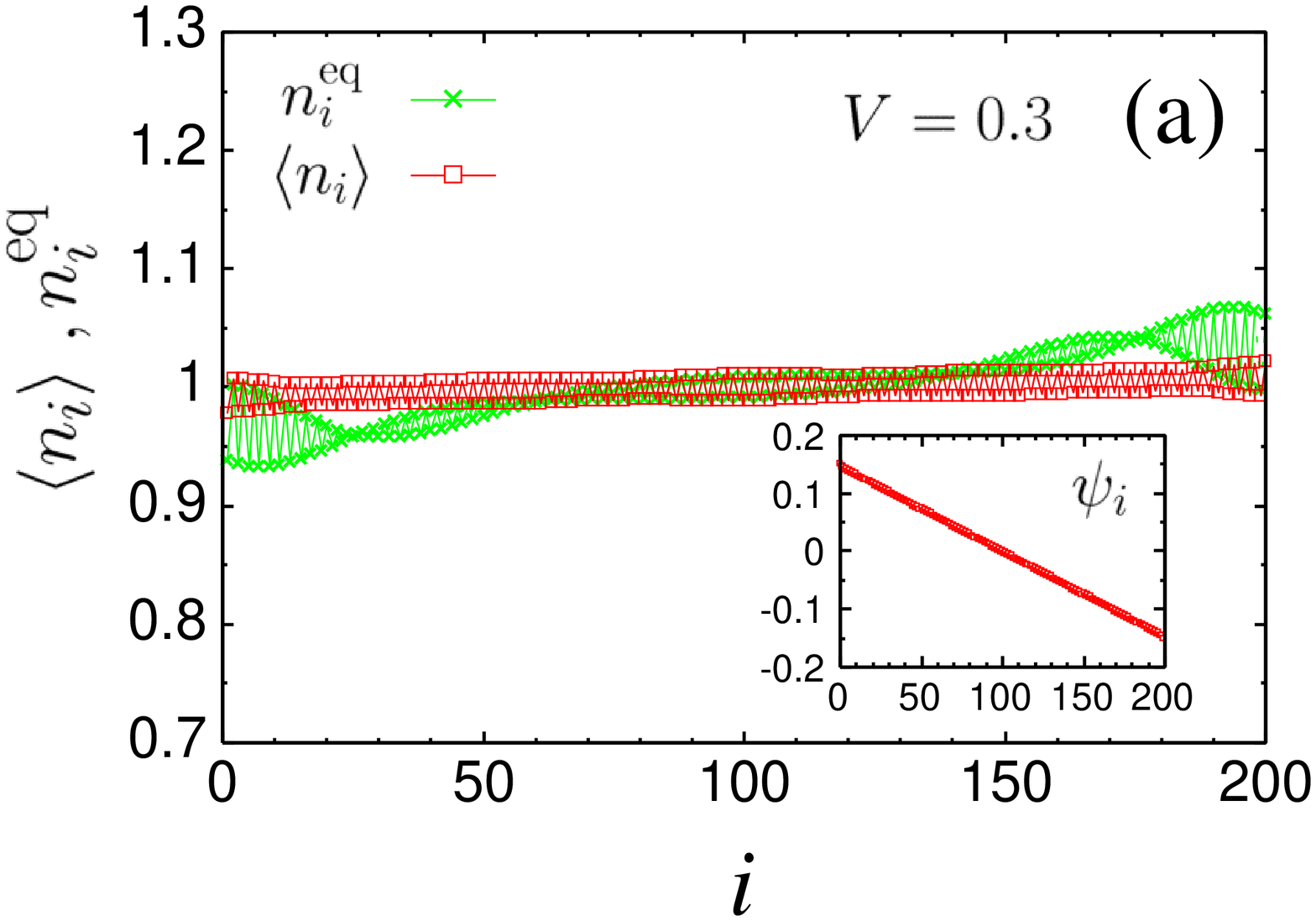}
\includegraphics[height=5.3cm]{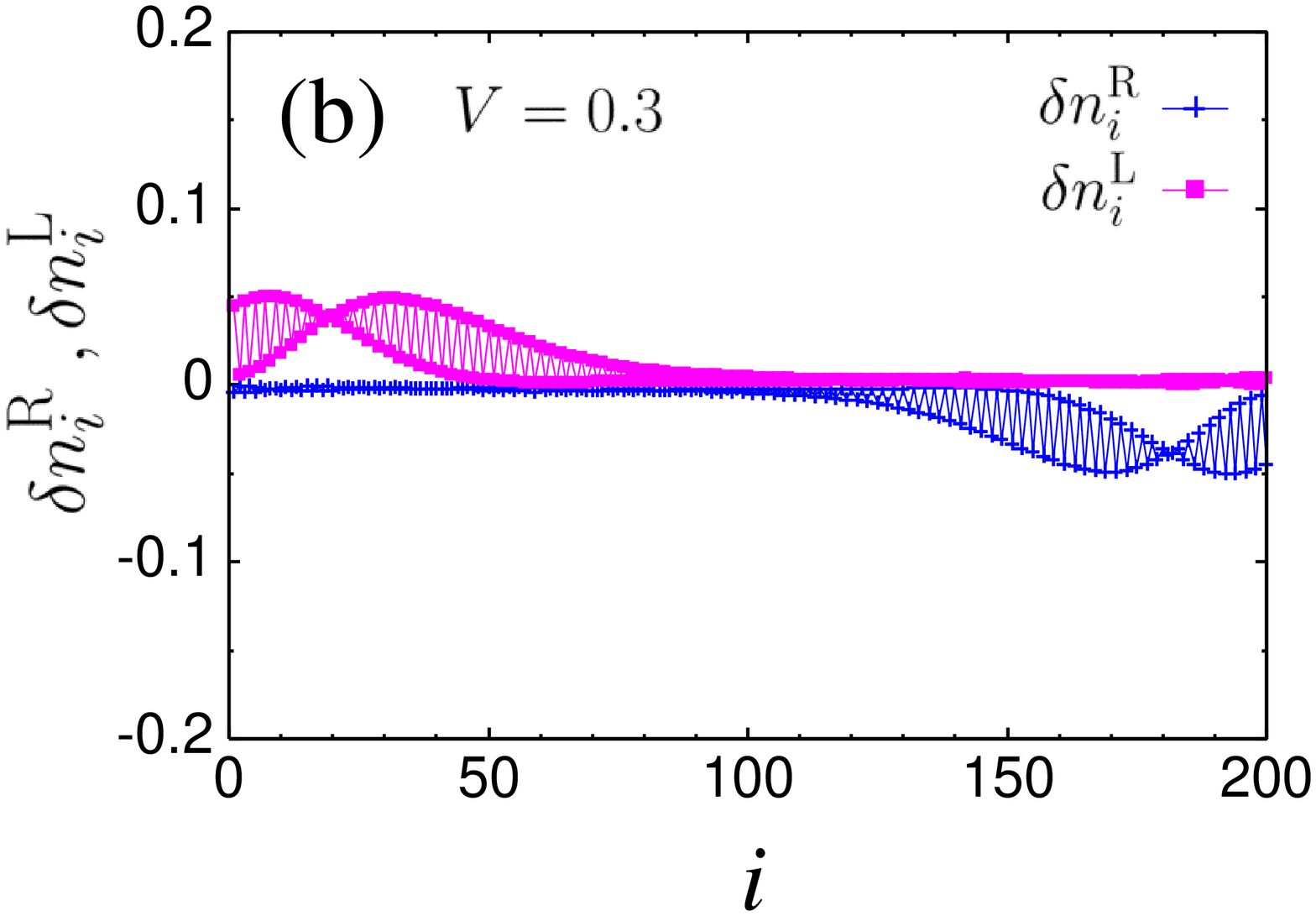}
\caption{(Color online) (a) Electron density, $\langle n_i\rangle$,
 ``equilibrium'' part, $n^{\rm eq}_i$, and (b) ``nonequilibrium'' parts
 $\delta n^{R}_i$ and $\delta n^{L}_i$ for $\delta t=0.025$, $L_C=200$,
 $U=0$, $V_p=0.1$, $\gamma_L =\gamma_R =0.1$, and $V=0.3$. The scalar
 potential $\psi_i$ is shown in the inset of (a).}
\label{fig:band_res}
\end{figure}
\begin{figure}
\includegraphics[height=5.3cm]{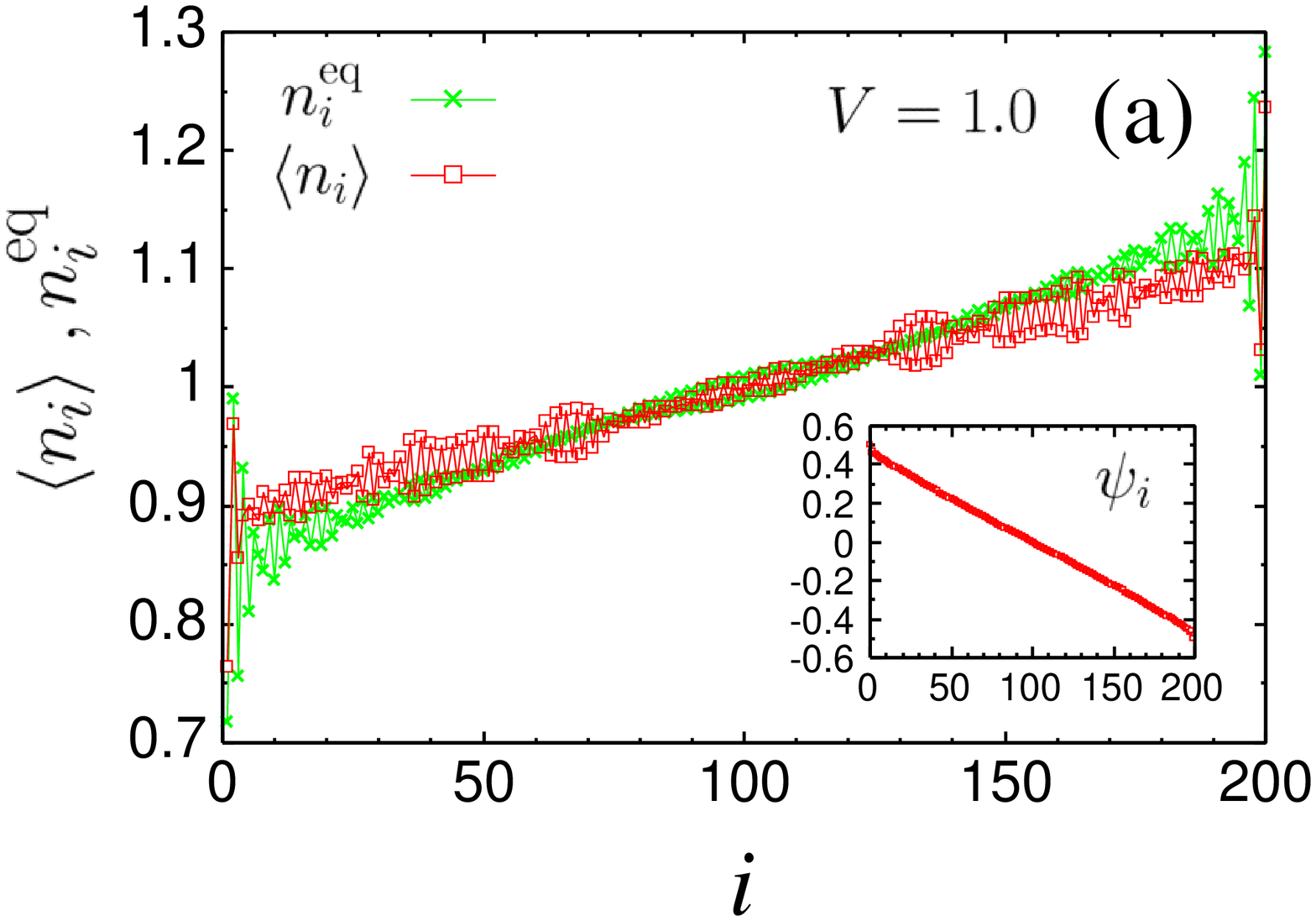}
\includegraphics[height=5.3cm]{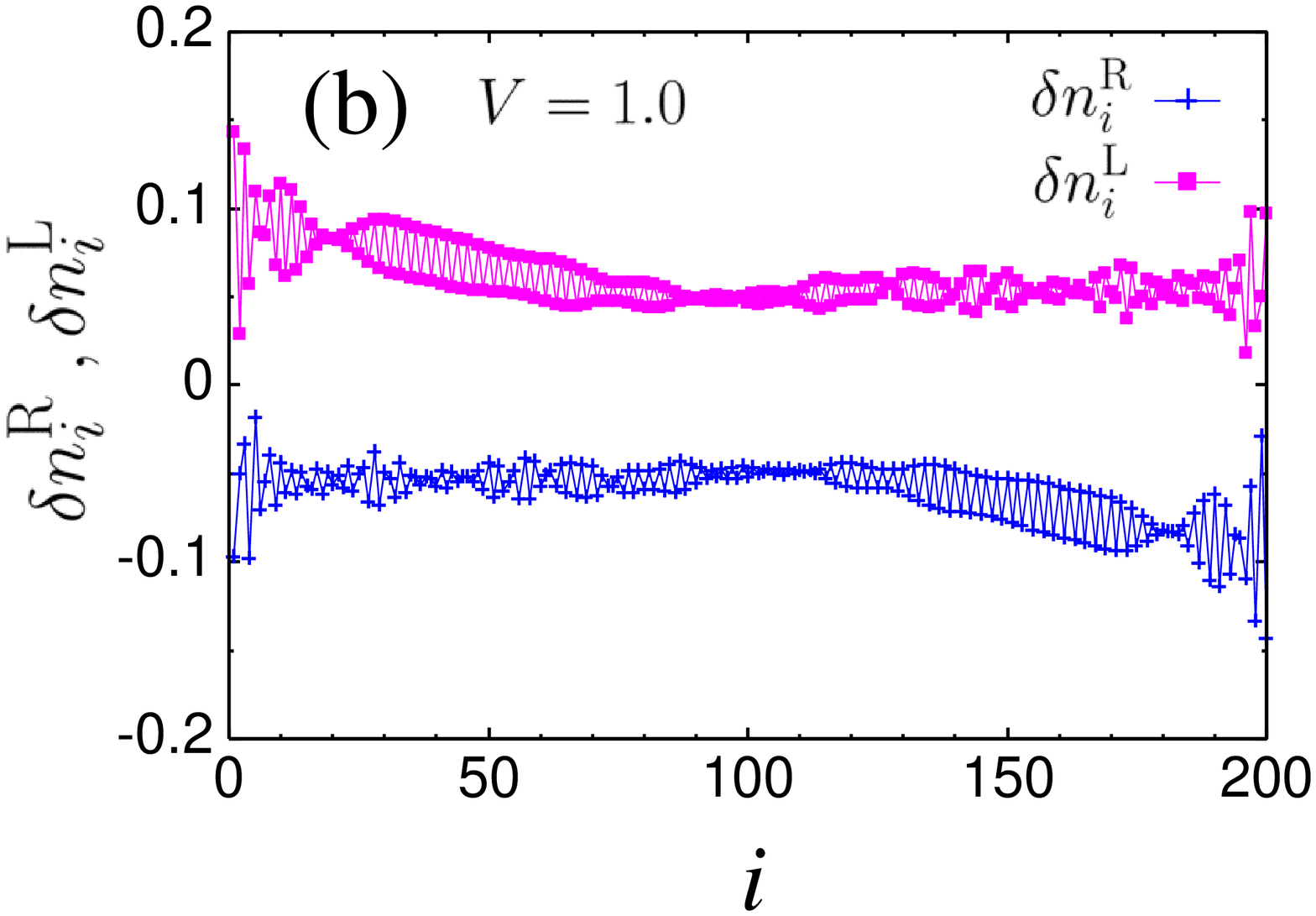}
\caption{(Color online) (a) Electron density, $\langle n_i\rangle$,
 ``equilibrium'' part, $n^{\rm eq}_i$, and (b) ``nonequilibrium'' parts
 $\delta n^{R}_i$ and $\delta n^{L}_i$ for $\delta t=0.025$, $L_C=200$,
 $U=0$, $V_p=0.1$, $\gamma_L = \gamma_R =0.1$, and $V=1.0$. The scalar
 potential $\psi_i$ is shown in the inset of (a).}
\label{fig:band_con}
\end{figure}
For finite voltages applied, the charge distributions in resistive and
conductive states for $\delta t=0.025$ are shown in
Figs. \ref{fig:band_res} and \ref{fig:band_con}, respectively. In
Fig. \ref{fig:band_res}(a), the electron density
$\langle n_{i}\rangle =\langle n_{i\uparrow}\rangle +\langle
n_{i\downarrow}\rangle$ and its ``equilibrium'' part 
$n^{\rm eq}_{i} = n^{\rm eq}_{i\uparrow} +n^{\rm eq}_{i\downarrow}$ 
for $V=0.3$ are shown. A $2k_F$ oscillation in the
charge distribution is induced by the boundaries.\cite{Yonemitsu_JPSJ09}
For all $i$, $\langle n_i\rangle$ is almost unity as
in the equilibrium case ($\langle n_i\rangle = n^{\rm eq}_i =1$ for
$V=0$). The electron density $\langle n_{i}\rangle$ is basically unchanged
by the bias voltage $V$ when $J$ is small. The scalar potential $\psi_i$
has a linear dependence on $i$ as shown in the inset of
Fig. \ref{fig:band_res}(a). 
This is because the long-range
interaction term in Eq. (\ref{eq:psi}) is small for 
$\langle n_i\rangle \sim 1$, so that $\psi_i$ is determined only by the
boundary conditions. 
The ``equilibrium'' part $n^{\rm eq}_i$,
on the other hand, deviates from unity near the left and right
electrodes, where the deviation is canceled by the ``nonequilibrium''
parts $\delta n^{\alpha}_i$, as shown in Fig. \ref{fig:band_res}(b). 
The quantity $\delta n^{\rm L}_i$ have nonnegative values for all $i$
because electrons come in from the left electrode. Although 
$\delta n^{\rm L}_i$ is large near the left
electrode, it decays as $i$ increases. On the other hand, 
$\delta n^{\rm R}_i$ have nonpositive values for all $i$ because
electrons go out to the right electrode. Note that 
$\delta n^{\rm R}_i=-\delta n^{\rm L}_{L_C+1-i}$ for $\gamma_L=\gamma_R$, no
work-function differences, and at half filling.\cite{Yonemitsu_JPSJ09}
The behaviors of $\delta n^{\rm L}_i$ and $\delta n^{\rm R}_i$ indicate that
electrons and holes hardly penetrate into the central part and the
resistive state is maintained. The current hardly flows through the central
part, because $J$ in
Eq. (\ref{eq:cur}) is determined by the difference between the density
modulation by the left electrode at the right boundary, 
$\delta n^{\rm L}_{i_R}$, and that by the right electrode at the left
boundary, $\delta n^{\rm R}_{i_L}$. Both terms $\delta n^{\rm L}_{i_R}$
and $\delta n^{\rm R}_{i_L}$ are vanishingly small, as shown in
Fig. \ref{fig:band_res}(b).

The charge distribution for $V=1.0$, where the system is conductive, is
qualitatively different from that for $V=0.3$ as shown in
Fig. \ref{fig:band_con}. The spatial dependences of 
$n^{\rm eq}_i$ and 
$\langle n_i\rangle$ are nearly the same. They increase almost linearly
from left to right except in the vicinities of the electrodes, where some
oscillatory structure appears.
The distributions
of $n^{\rm eq}_i$ and $\langle n_i\rangle$ are understood by that of
$\psi_i$ shown in the inset of Fig. \ref{fig:band_con}(a).
The electron density is higher (lower) on the right (left) half where
$\psi_i$ is low (high). This behavior is caused by the electrons that
move through the system in the conductive phase. 
The profile of
$\psi_i$ shows almost a linear dependence on $i$ although a small deviation
from the linearity near the electrodes is visible in contrast to the
resistive phase, which is because the charge redistribution 
($\langle n_i\rangle\neq 1$) is easier in the conductive phase.
As for the ``nonequilibrium'' parts of the
density shown in Fig. 4(b), $\delta n^{\rm L}_i$ have
positive values for all $i$, while $\delta n^{\rm R}_i$ are negative
for all $i$ because the electrons come in from the left
electrode and go out to the right electrode. A finite current flows
through the central part: $\delta n^{\rm L}_{i_R}$ and 
$\delta n^{\rm R}_{i_L}$ are finite and have the opposite signs.
\begin{figure}
\includegraphics[height=5.7cm]{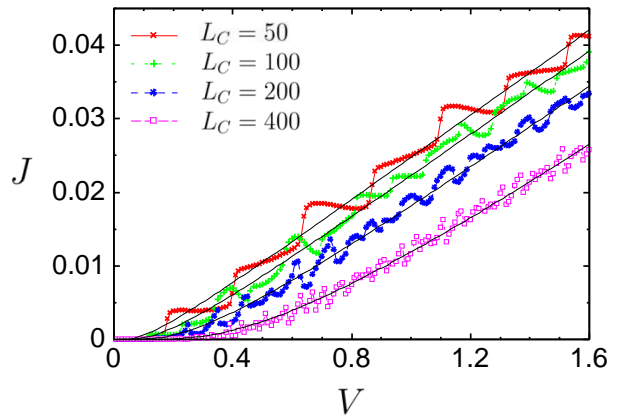}
\caption{(Color online) $I$-$V$ characteristics of one-dimensional band
 insulators for several values of $L_C$ with $\delta t=0.025$, $U=0$,
 $V_p=0.1$, and $\gamma_L =\gamma_R =0.1$. The solid lines show the
 function $J=aVe^{-V_{\rm th}/V}$ which fits to the results.}
\label{fig:band_iv2}
\end{figure}

\begin{figure}
\includegraphics[height=5.7cm]{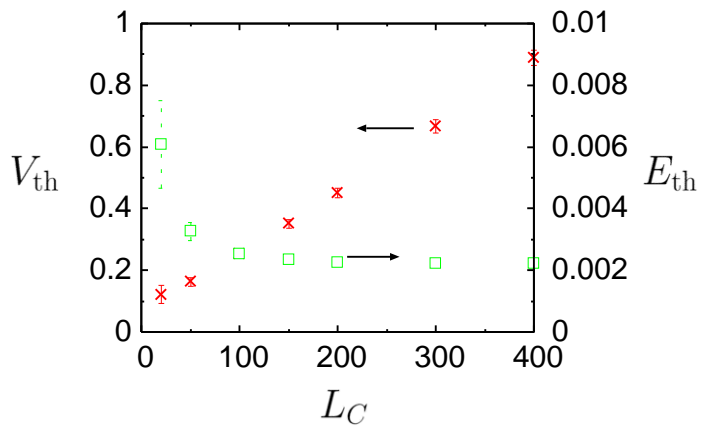}
\caption{(Color online) Dependence of the threshold bias voltage $V_{\rm
 th}$ and the threshold electric field $E_{\rm th}$ on the size of the
 central part $L_C$. The other parameters are the same as in
 Fig. \ref{fig:band_iv2}. The error bars in the fitting are also shown.}
\label{fig:band_cross}
\end{figure}
Next, we discuss the breakdown mechanism of band insulators.
In Fig. \ref{fig:band_iv2}, we show the $I$-$V$ curves for
different sizes of the central parts $L_C$ with $\delta t=0.025$. To the
numerical results, the function, 
\begin{equation}
J=aVe^{-V_{\rm th}/V},
\label{eq:lz_iv}
\end{equation}
is well fitted, where $a$ and $V_{\rm th}$ are parameters. This
expression originates from the LZ tunneling mechanism through which
the insulator breaks down with the threshold voltage 
$V_{\rm th}$.\cite{Oka_PRL05} For $V<V_{\rm th}$, the current $J$ is
exponentially
suppressed due to the energy gap, while it increases linearly for
$V>V_{\rm th}$.
When the central part is large, the fitting works
well as shown in Fig. \ref{fig:band_iv2}, so that the breakdown is
consistent with the LZ tunneling picture, although there exist fine
structures in the $I$-$V$ characteristics which come from the
discreteness of the energy spectrum of the central part. As $L_C$
decreases, the structure becomes more prominent. For $L_C=50$, for
example, a deviation from the fitting curve due to the
stepwise structure becomes large, which indicates the LZ mechanism is no
longer applicable to small-$L_C$ systems.

Figure \ref{fig:band_cross} shows $V_{\rm th}$ determined by fitting
Eq. (\ref{eq:lz_iv}) to the data for each $L_C$, together with the
corresponding electric field $E_{\rm th}\equiv V_{\rm th}/L_C$. For
large $L_C$,
$V_{\rm th}$ is proportional to $L_C$, so that $E_{\rm th}$ becomes a
constant. In band insulators, the LZ breakdown is known
to be induced by the applied electric
field.\cite{Landau_SOW32,Zener_PRSL32} Since the
one-particle picture holds in band insulators, this breakdown can be
analyzed as a usual interband tunneling problem and the threshold
electric field becomes $E_{\rm th}\propto \Delta^2/W$.\cite{Ziman} The
breakdown occurs when the energy gain by displacing an electron with
charge $-e$ in an electric field $E$ by the distance $\xi=W/\Delta$,
$eE\xi$, overcomes the energy gap $\Delta$. Here, $W\simeq 4$ is the
bandwidth and $\xi\sim 40$. We have obtained the threshold 
$E_{\rm th}=0.0022$, which is comparable
with the value obtained by the LZ
formula,\cite{Landau_SOW32,Zener_PRSL32} $(\Delta/2)^2/v=0.00125$ with
$v=2$. In short, the threshold is governed by the electric field. 

When the central part is small, the fitting to the $I$-$V$ curve becomes
worse because the finite-size effect becomes severe. The LZ mechanism is
not suitable for understanding this breakdown. In this case, another
mechanism, in which the threshold is determined by the bias voltage, is
more appropriate for the following reason. As $L_C$ decreases, it
eventually becomes smaller than the correlation length $\xi$. The
tunneling occurs when the energy gain by displacing an electron by
the distance $L_C$, $eEL_C$, overcomes the energy gap $\Delta$.
This indicates that the mechanism of the
breakdown continuously changes around $L_C\sim \xi$ as a function of
$L_C$. When $V$ exceeds
$\Delta$, some energy levels of the central part come in between $\mu_L$
and $\mu_R$. For $L_C<\xi$, the wave functions do not fully decay in the
system: the electron injected from the left electrode
with energy higher than $\Delta/2$ can reach the right electrode through
these levels so that the current flows. This can be clearly seen in
Fig. \ref{fig:band_iv2} for $L_C=50$ where the gap is
$\Delta\sim 0.2$ due to the finite-size effect. The $I$-$V$ curve shows
an abrupt increase at $V\sim \Delta$ because $\mu_L$ exceeds
the lowest unoccupied energy level of the central part. Each stepwise
increase in the $I$-$V$ characteristics corresponds to the increase in
the number of energy levels located between $\mu_L$ and $\mu_R$. 

We have numerically confirmed that the results are qualitatively
unchanged even if the long-range Coulomb interaction strength $V_p$ and
the system-electrode coupling strength $\gamma_L$ ($=\gamma_R$) are
varied. Thus, the threshold shows a crossover as a function of $L_C$. 
When $L_C>\xi$, the LZ-type breakdown occurs and the threshold is
governed by the electric field. For $L_C<\xi$, on the other hand, the
current flows when $V$ exceeds the energy gap $\Delta$.

It is noted that the spatial dependence of $\psi_i$ is important for
the realization of the field-induced breakdown as well as the spatial
modulation of the wave functions as discussed below. For small $V$,
$\psi_i$ has a linear dependence on $i$ throughout the central part
because the electrons are localized, $\langle n_i\rangle\simeq 1$, so
that the effect of the long-range interaction on $\psi_i$ is small. When
the system is conductive, the charge redistribution occurs near the
interfaces where a small deviation from the linearity is seen in
$\psi_i$. This charge redistribution weakens the electric field on the
sites away from the interfaces in the central part.
For comparison, we show in the Appendix the $I$-$V$ characteristics that
are
obtained by artificially setting $\psi_i=0$ for all $i$. This
corresponds to a hypothetical case where a sufficiently large charge
redistribution occurs near the interfaces. There is no electric field in
the central part: a voltage drop occurs only at the interfaces. 
In this extreme case, we obtain the threshold bias voltage
$V_{\rm th}\sim \Delta$ regardless of the length $L_C$, which is in
contrast to the LZ-type behavior for large $L_C$ in
Fig. \ref{fig:band_cross}.
Such a situation never occurs in our calculations with
$\psi_i$ and for realistic parameters. As we will discuss in the next
section and also in the Appendix, the effect of the spatial profile of
$\psi_i$ on the breakdown mechanism of Mott insulators is
basically the same as in the case of band insulators. Thus, the model
without $\psi_i$ is inappropriate for realistic insulators.  
\begin{figure}
\includegraphics[height=6.6cm]{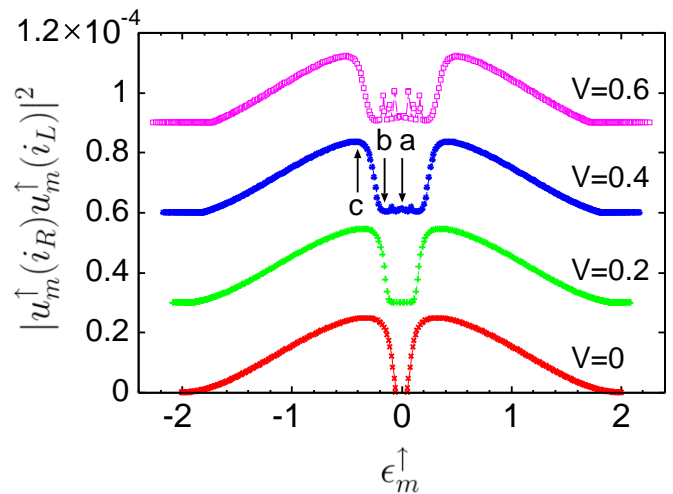}
\caption{(Color online) $|u^{\sigma}_m(i_R)u^{\sigma}_m(i_L)|^2$ with
 $\sigma =\uparrow$ plotted against $\epsilon^{\uparrow}_m$ for $V=0$,
 $0.2$, $0.4$, and $0.6$ in the case of $\delta t=0.025$, $L_C=400$,
 $U=0$, $V_p=0.1$, and $\gamma_L=\gamma_R=0.1$. For $V=0.2,0.4$,
 and $0.6$, the lines are shifted upward by $0.3$, $0.6$, and $0.9$,
 respectively.}
\label{fig:band_uiRuiL}
\end{figure}

In discussing the breakdown for $L_C\gg \xi$, the spatial dependences of
the wave functions $u^{\sigma}_m(i)$ are crucial as explained above. 
In Fig. \ref{fig:band_uiRuiL}, we show
$|u^{\uparrow}_m(i_R)u^{\uparrow}_m(i_L)|^2$ as a
function of the real part of the one-particle energy 
$\epsilon^{\uparrow}_m$ for several values of $V$ in the case of
$L_C=400$. The behavior of this
quantity for $\sigma=\downarrow$ is the same. It shows whether a given
one-particle state contributes to the current $J$. Note that 
$J$ is obtained by integrating $|[G^{r}_{\sigma}(\epsilon)]_{i_Li_R}|^2$
over $\mu_R<\epsilon<\mu_L$. Since $\epsilon -\epsilon^{\sigma}_m$
appears in the denominator for
$[G^{r}_{\sigma}(\epsilon)]_{ij}$ [Eq. (\ref{eq:gr})], 
the one-particle state $m$ with finite
$|u^{\sigma}_m(i_R)u^{\sigma}_m(i_L)|^2$ in the interval 
$-V/2=\mu_R<\epsilon^{\sigma}_m<\mu_L=V/2$ gives a large contribution
to $J$. This quantity directly shows whether the
one-particle state $m$ is localized or delocalized because it comes from
the product of the amplitudes of the wave
function at the two interfaces $i_L$ and $i_R$. If
$|u^{\sigma}_m(i_R)u^{\sigma}_m(i_L)|^2$ is large, the state has finite
amplitudes at both sides of the central part, so that it is
delocalized. If $|u^{\sigma}_m(i_R)u^{\sigma}_m(i_L)|^2$ is small, on
the other hand, the state has a small amplitude at either of the
interfaces.

For $V=0$, $|u^{\uparrow}_m(i_R)u^{\uparrow}_m(i_L)|^2$ shows two bands
that correspond to the conduction and valence bands in the band
insulator. Because of the energy gap, no state
exists in the region $-\Delta/2<\epsilon^{\sigma}_m<\Delta/2$ for $V=0$
so that
the current does not flow at least for $V<\Delta$. When $V=0.2$, several
states appear in the region $-\Delta/2<\epsilon^{\uparrow}_m<\Delta/2$,
corresponding to the leakage of one-particle states from the electrodes
to the central part. However, these states do not contribute to the
current because
their $|u^{\uparrow}_m(i_R)u^{\uparrow}_m(i_L)|^2$ are vanishingly
small as shown in Fig. \ref{fig:band_uiRuiL}. Note that the
line for each $V\neq 0$ is
shifted upward by $\frac{3}{2}V$. For $V=0.4$, the number of
states around $\epsilon^{\uparrow}_m =0$ with small
$|u^{\uparrow}_m(i_R)u^{\uparrow}_m(i_L)|^2$ increases. 
The energy range where these localized
states appear becomes wider as $V$ increases. Consequently, the
delocalized states that contribute to the current depart from the region
$\mu_R<\epsilon<\mu_L$. Therefore, the current does not flow even if $V$
barely exceeds the gap. As we show in the Appendix,
the localized states do not appear if we set $\psi_i=0$ for all $i$. It
is crucial to take the spatial dependence of $\psi_i$ into
account to obtain the modulation of the wave functions. 
\begin{figure}
\includegraphics[height=9.0cm]{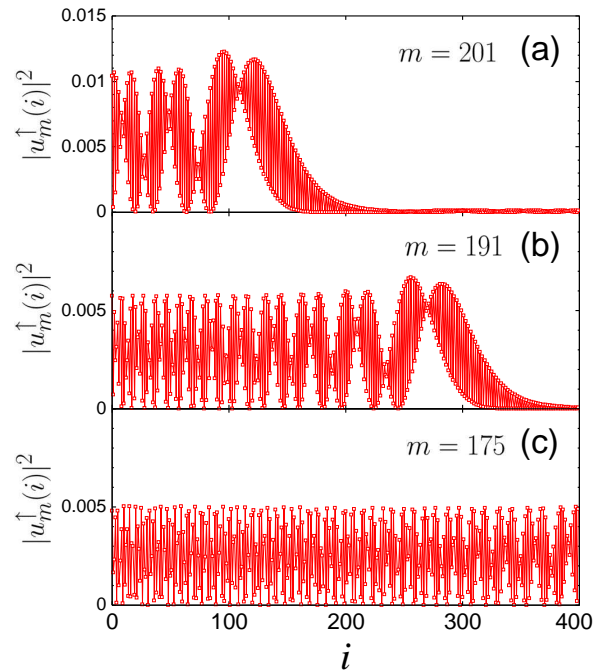}
\caption{(Color online) Spatial dependence of one-particle states
 $|u^{\uparrow}_m(i)|^2$ for (a) $m=201$, (b) $m=191$, and (c) $m=175$
 in the case of $V=0.4$. The other parameters are the same as in
 Fig. \ref{fig:band_uiRuiL}. The corresponding $\epsilon^{\uparrow}_m$
 are indicated by the arrows in Fig. \ref{fig:band_uiRuiL}.}
\label{fig:band_wf}
\end{figure}

Figure \ref{fig:band_wf} shows the spatial dependences of the squares of
the absolute values of the wave functions $|u^{\uparrow}_m(i)|^2$ for
$V=0.4$ and several $m$ whose $\epsilon^{\uparrow}_m$ are located at the
positions indicated by the arrows in Fig. \ref{fig:band_uiRuiL}. Here
$m$ is so labeled that
$\epsilon^{\sigma}_m<\epsilon^{\sigma}_{m^{\prime}}$ for
$1\leq m<m^{\prime}\leq L_C=400$.
Figure \ref{fig:band_wf}(a) shows the one-particle state in the lower
band with $\epsilon^{\uparrow}_m=0.0348$ ($m=201$) which is inside the
gap for $V=0$. This state is localized on the left half of the central
part. The reason is as follows. The scalar potential is high (low) near
the left (right)
electrode. Within each of the conduction and valence bands, the state
whose weight is large near the left (right) electrode has a higher
(lower) energy than others. In the present case, 
the indexes $m$ for the valence band are $m=206$, $203$, $201$, 
$199,\ldots$, $3$, $2$, $1$,
while those for the conduction band are $m=195$, $198$, $200$, 
$202,\ldots$, $398$, $399$, $400$. As $m$ is
lowered, the wave function of the one-particle state is generally
extended to a wider region and its largest amplitude is shifted to the
right, as shown in Fig. \ref{fig:band_wf}(b) for the case of 
$\epsilon^{\uparrow}_m=-0.157$ ($m=191$). 
As $m$ is lowered further, e.g., for $\epsilon^{\uparrow}_m=-0.400$
($m=175$) in Fig. \ref{fig:band_wf}(c), the one-particle state is
delocalized to reach the right electrode. Then, its wave function has
large amplitudes near both electrodes.

\begin{figure}
\includegraphics[height=8.0cm]{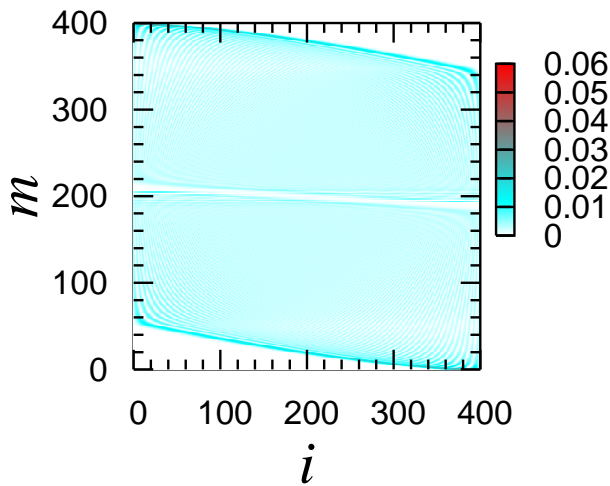}
\caption{(Color online) Contour map of one-particle states
 $|u^{\uparrow}_m(i)|^2$ in $(i,m)$ plane for $V=0.4$. The other
 parameters are the same as in Fig. \ref{fig:band_uiRuiL}.}
\label{fig:band_map}
\end{figure}

In order to overview the behaviors of the one-particle states, we show
the contour map of one-particle states $|u^{\uparrow}_m(i)|^2$ on the
$(i,m)$ plane for $V=0.4$ in Fig. \ref{fig:band_map}. For $V=0.4$, the
states $190\lesssim m\lesssim 200$ are localized near the left electrode,
and the states below are delocalized. At the bottom of the lower band,
the states are localized near the right electrode because of the low
scalar potential near the right electrode. 

\subsection{Mott insulators}
\begin{figure}
\includegraphics[height=5.7cm]{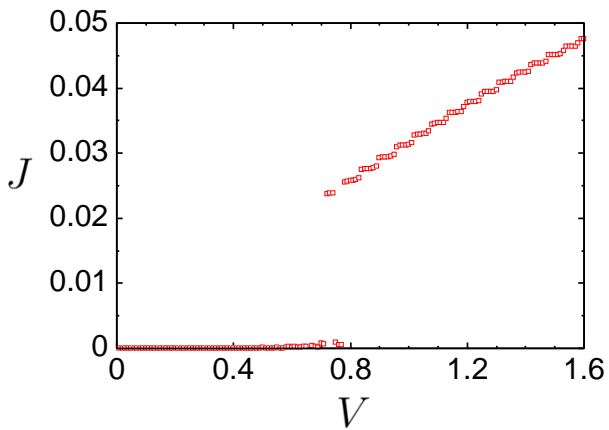}
\caption{(Color online) $I$-$V$ characteristics of one-dimensional
 Hubbard model for $\delta t=0$, $L_C=200$, $U=1.5$, $V_p=0.1$, and 
$\gamma_L =\gamma_R =0.1$.}
\label{fig:mott_iv}
\end{figure}

In this section, we consider the case where the central part is
described by the Hubbard model. Figure \ref{fig:mott_iv} shows the
$I$-$V$ characteristics for $\delta t=0$, $L_C=200$, $U=1.5$, $V_p=0.1$,
and $\gamma_L =\gamma_R = 0.1$. When $V=0$, the system is an
antiferromagnetic insulator owing to the Hartree-Fock approximation. The
energy gap $\Delta$ is then 0.25. 
In the previous studies on the $I$-$V$ characteristics of
metal-Mott-insulator interfaces,\cite{Yonemitsu_JPSJ05,Yonemitsu_PRB07}
the current was
calculated by solving the time-dependent Sch$\ddot{\rm o}$dinger
equation. It is argued that the results obtained
by the time-dependent Hartree-Fock approximation for the
electron-electron interaction are consistent with those obtained by
exact many-electron wave
functions on small systems. For example, the suppression of
rectification at metal-Mott-insulator interfaces is described by both
methods.\cite{Yonemitsu_PRB07} Although the present time-independent
Hartree-Fock approximation is worse, we expect the present approach 
captures the essential features of nonequilibrium steady states under
the bias voltage. As we increase $V$, the current begins to flow at
$V_{\rm th}\sim 0.8$. The breakdown becomes a first-order transition
due to the Hartree-Fock approximation, which is in contrast to the case
of band insulators in the previous section. 
\begin{figure}
\includegraphics[height=5.3cm]{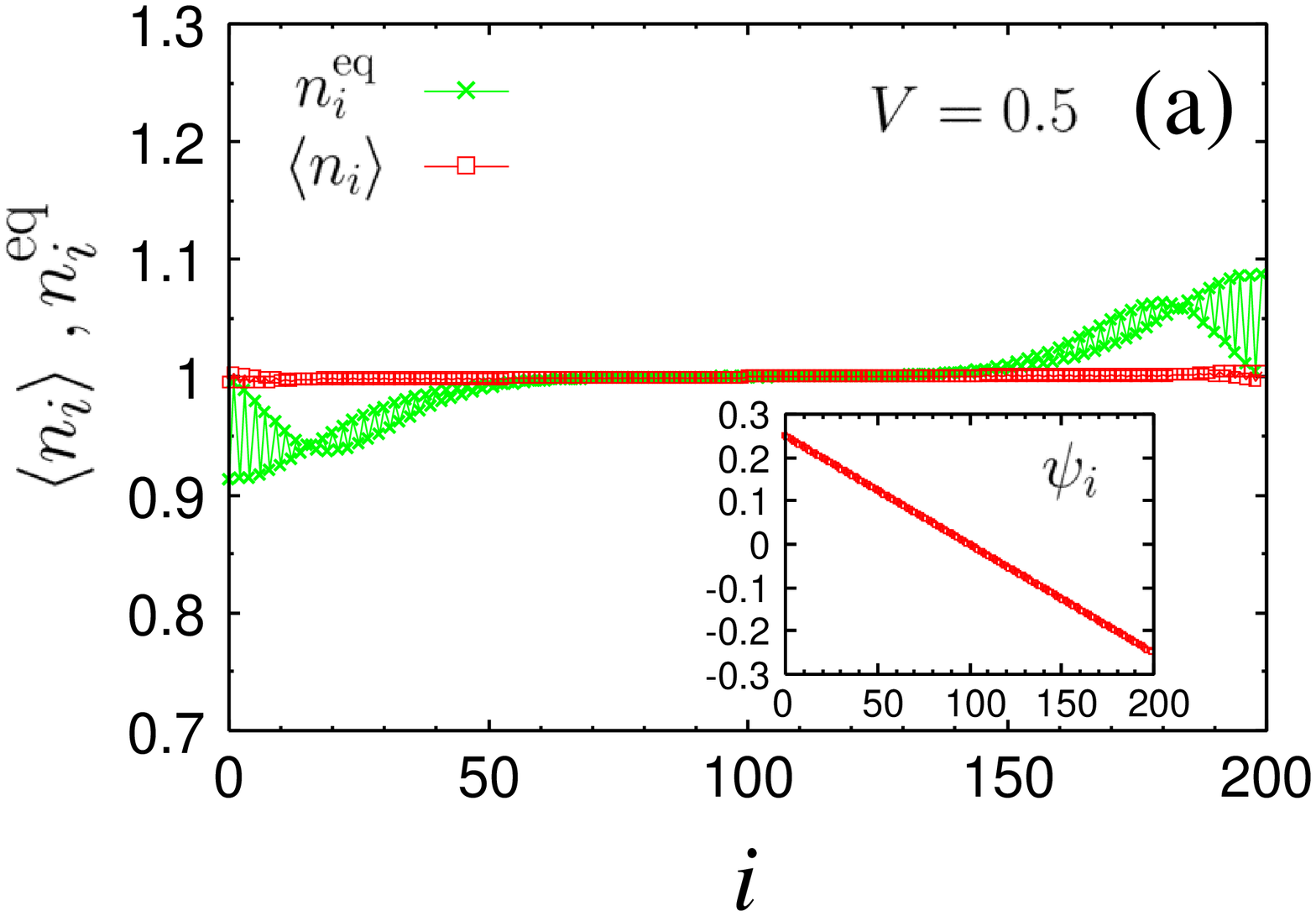}
\includegraphics[height=5.3cm]{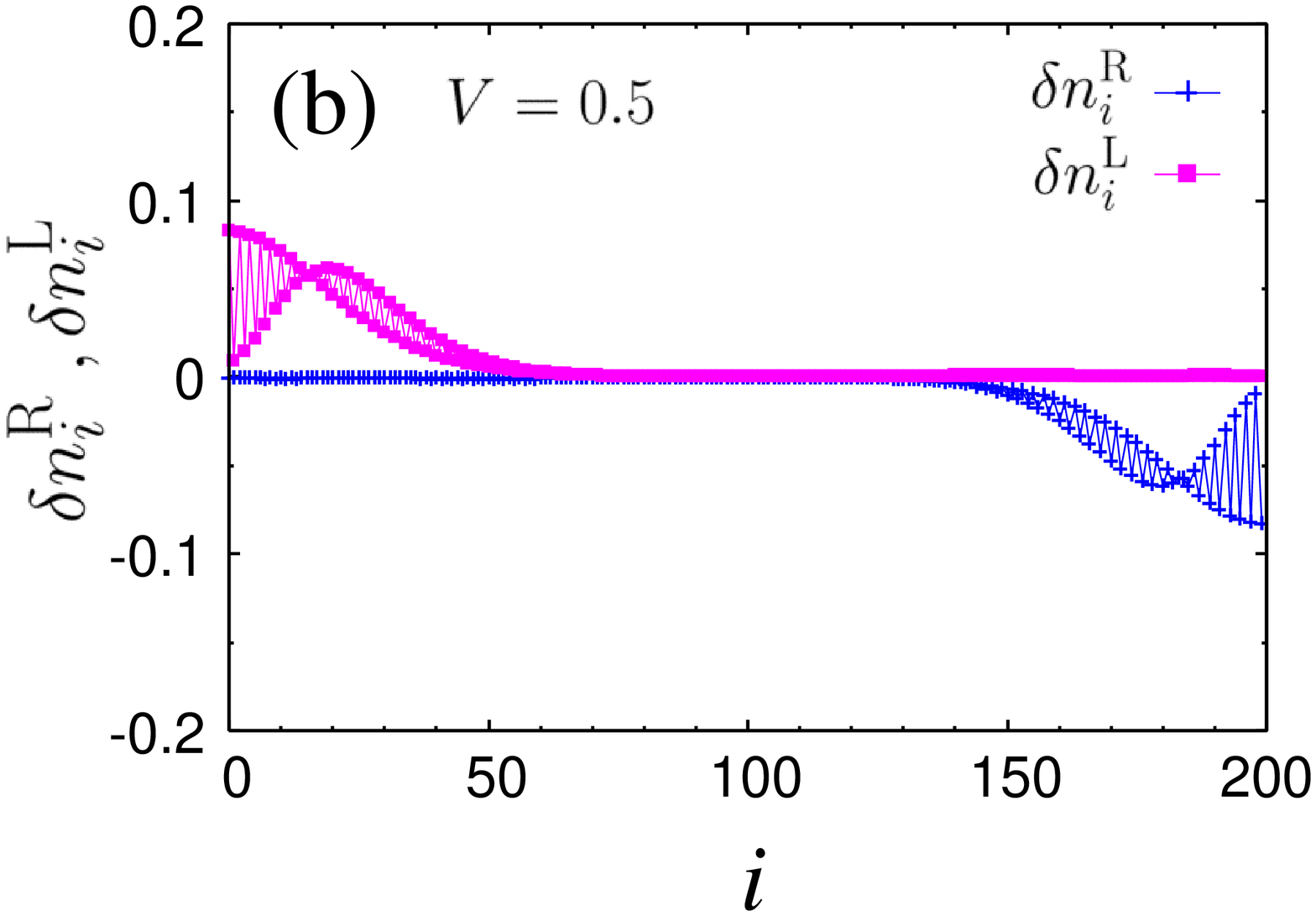}
\caption{(Color online) (a) Electron density, $\langle n_i\rangle$,
 ``equilibrium'' part, $n^{\rm eq}_i$, and (b) ``nonequilibrium'' parts
 $\delta n^{R}_i$ and $\delta n^{L}_i$ for $\delta t=0$, $L_C=200$,
 $U=1.5$, $V_p=0.1$, $\gamma_L =\gamma_R =0.1$, and $V=0.5$. The scalar
 potential $\psi_i$ is shown in the inset of (a).}
\label{fig:mott_res}
\end{figure}
\begin{figure}
\includegraphics[height=5.3cm]{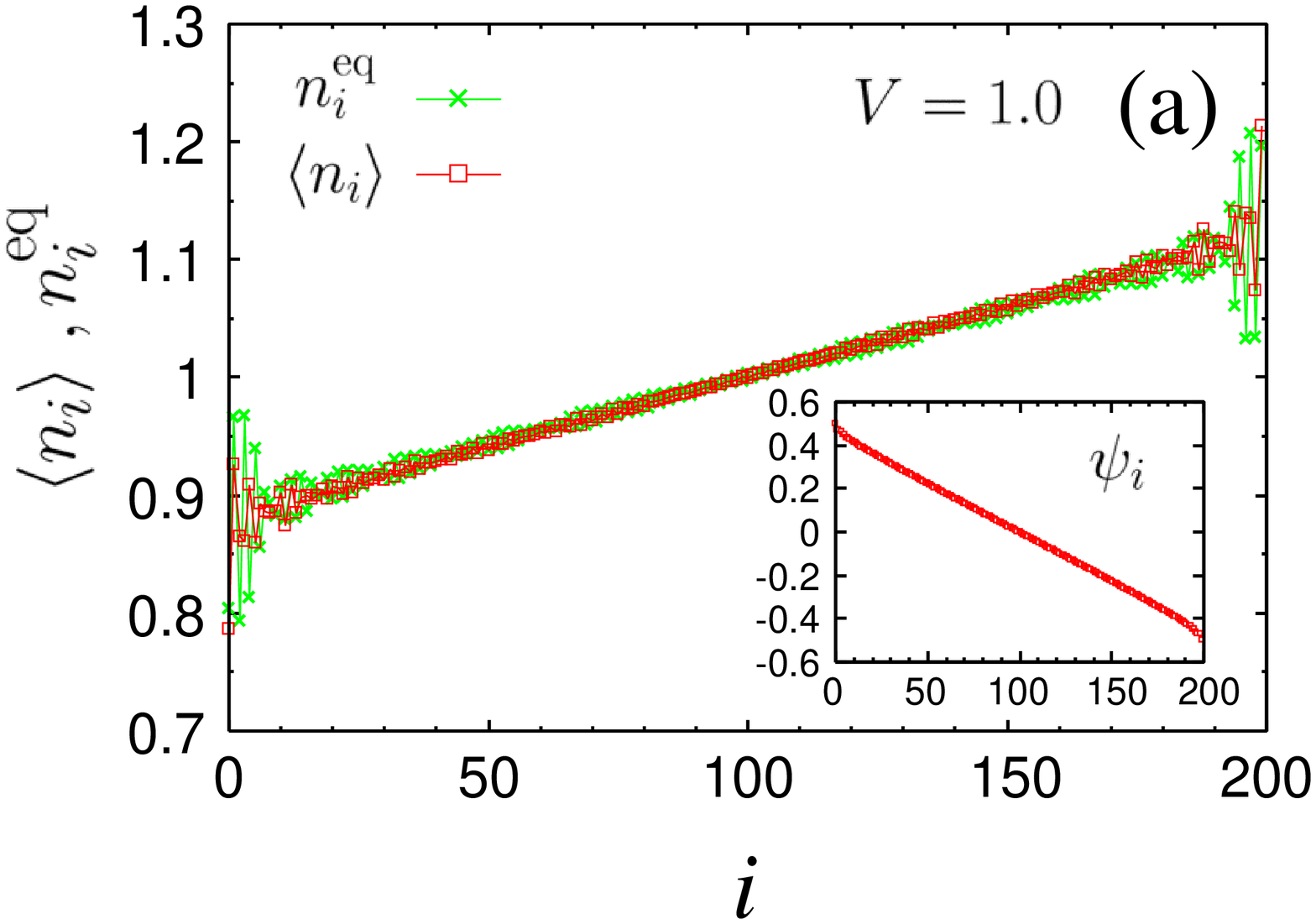}
\includegraphics[height=5.3cm]{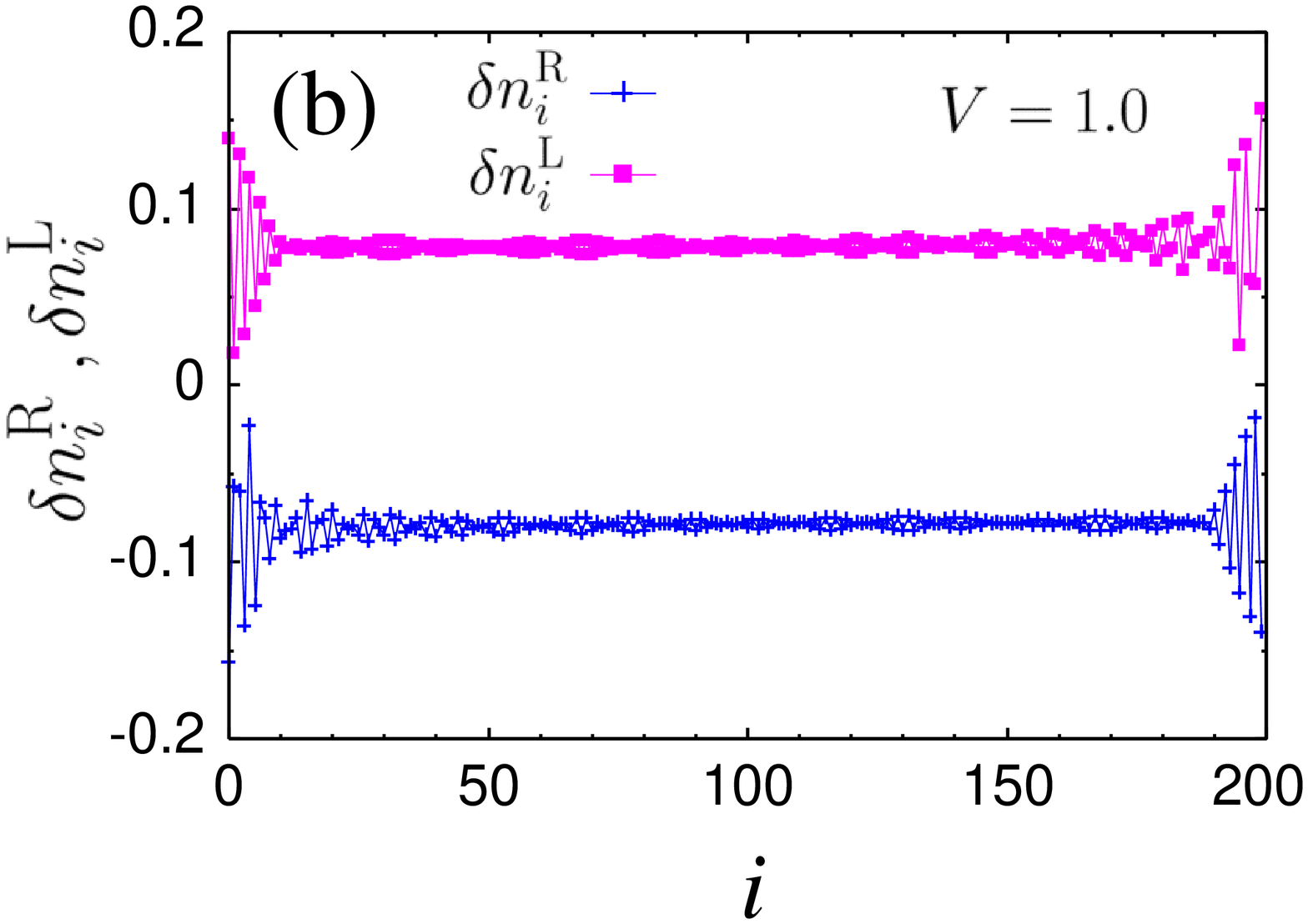}
\caption{(Color online) (a) Electron density, $\langle n_i\rangle$,
 ``equilibrium'' part, $n^{\rm eq}_i$, and (b) ``nonequilibrium'' parts
 $\delta n^{R}_i$ and $\delta n^{L}_i$ for $\delta t=0$, $L_C=200$,
 $U=1.5$, $V_p=0.1$, $\gamma_L = \gamma_R =0.1$, and $V=1.0$. The scalar
 potential $\psi_i$ is shown in the inset of (a).}
\label{fig:mott_con}
\end{figure}

The charge distributions in resistive and conductive phases at finite
$V$ are shown in Figs. \ref{fig:mott_res} and \ref{fig:mott_con},
respectively. Their overall features are similar to those in
band insulators. For all $i$ and $V<V_{\rm th}$, the electron density 
$\langle n_i\rangle$ is almost unity, which is basically the same as in
the equilibrium case ($\langle n_i\rangle = n^{\rm eq}_i =1$ for $V=0$). 
The scalar potential $\psi_i$ has a linear dependence on $i$ as shown in
the inset of Fig. \ref{fig:mott_res}(a). As for the ``nonequilibrium''
parts, $\delta n^{\rm L}_i$ 
($\delta n^{\rm R}_i$) is large near the left (right) electrode and
decays as $i$ increases (decreases) [Fig. 11(b)]. Electrons and holes do
not penetrate into the central part so that the current does not flow.

For $V=1.0>V_{\rm th}$, the spatial dependences of $n^{\rm eq}_i$ and 
$\langle n_i\rangle$ are shown in Fig. \ref{fig:mott_con}(a). They
increase almost linearly from left to right. This reflects the profile
of the scalar potential $\psi_i$ that is higher (lower) on the left
(right) half. The ``nonequilibrium'' parts of the densities, 
$\delta n^{\rm L}_i$ and $\delta n^{\rm R}_i$ are extended over the
whole system with small spatial dependence
[Fig. 12(b)],\cite{Yonemitsu_JPSJ09} which
is in contrast to the resistive phase. Since  $\delta n^{\rm L}_{i_R}$
and $\delta n^{\rm R}_{i_L}$ are finite with opposite signs, a finite
current flows through the central part.

\begin{figure}
\includegraphics[height=5.5cm]{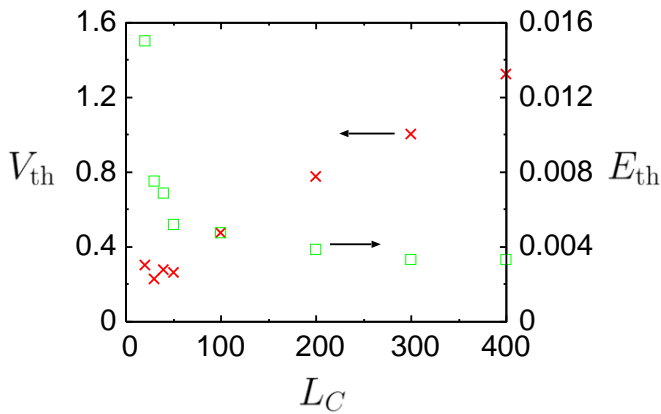}
\caption{(Color online) Dependence of the threshold bias voltage 
$V_{\rm th}$ and the threshold electric field $E_{\rm th}$ on the size
 of the central part $L_C$. The other parameters are the same as in
 Fig. \ref{fig:mott_iv}.}
\label{fig:mott_cross}
\end{figure}

In Fig. \ref{fig:mott_cross}, we show the threshold bias voltage 
$V_{\rm th}$ and the corresponding electric field 
$E_{\rm th}=V_{\rm th}/L_C$ for the first-order transition as a function
of $L_C$. 
When the central part is small, i.e., $L_C\lesssim 50$, $V_{\rm th}$
is almost a constant near the energy gap $\Delta\simeq 0.25$. For small
$L_C$, the electron injected from the left electrode with energy higher
than $\Delta/2$ can reach the right electrode since the correlation
length $\xi =W/\Delta\simeq 16$ is comparable to $L_C$. Thus, the
threshold for $L_C<\xi$ is determined by the bias voltage. For 
$L_C\gg \xi$, on
the other hand, $V_{\rm th}$ is proportional to $L_C$, so that the
threshold is governed by the electric field. 
In recent theoretical studies,\cite{Oka_PRL05} Oka and Aoki have
proposed that the LZ breakdown occurs also in Mott insulators by
applying the time-dependent density-matrix-renormalization-group method
to the one-dimensional Hubbard model
under an electric field with open boundary condition. In our
calculations, the scalar potential $\psi_i$ is linearly increasing with
$i$ in the resistive phase as shown in Fig. \ref{fig:mott_res} (a),
which means that the electrons feel a uniform electric field in the
central part. Therefore, our model describes the LZ breakdown as in the
open Hubbard chain as long as the electrodes do not affect the nature of
the breakdown for $L_C\gg \xi$.
In fact, the threshold $E_{\rm th}$ is about 0.0033, which is
comparable to the LZ
value,\cite{Oka_PRL05} $(\Delta/2)^2/v=0.0078$, with $v=2$. 
Thus, the threshold shows a crossover as a function of $L_C$ as in the
case of band insulators.

The LZ breakdown is explained as before by comparing the charge gap
$\Delta$ and the work which is done by the electric field on an electron
moving over the correlation length $\xi$. If the work $e E\xi$ exceeds
$\Delta$, the electron in the lower band may go over to the upper band
so that the current
flows. According to the results by Oka and Aoki,\cite{Oka_PRL05} this
consideration is applicable to Mott insulators where the correlation
effects are important. Thus, we expect that the results obtained
by the Hartree-Fock approximation are qualitatively unchanged even if we
take account of the electron correlation. It is well known that the
Hartree-Fock theory overestimates the charge gap $\Delta$. It predicts
the antiferromagnetic spin ordering which is actually destroyed if
quantum fluctuations are appropriately taken into account. However, the
overestimated $\Delta$ will alter the
threshold only quantitatively. We also expect that the spin ordering
does not essentially affect the breakdown itself since only the charge
degrees of freedom are relevant to the mechanism. We note that the
time-dependent density-matrix-renormalization-group
method has been applied to a Mott insulator with electrodes very
recently,\cite{Heidrich_Condmatt} and that the $I$-$V$ characteristics
have been consistently explained by
the LZ tunneling mechanism. However, it is also shown that physical
quantities such as the spin structure factor and the double occupancy do
not reach a stationary state in the accessible time window. Therefore, a
direct description of nonequilibrium steady states as in the present
study is important to deepen our understanding of the breakdown.

Since the crossover behavior is obtained for both band and Mott
insulators, the electron-electron interaction is not responsible for the
phenomenon. We emphasize that the spatial profile of $\psi_i$ is important
for the realization of the LZ breakdown. In the Appendix, this is
demonstrated for the Mott insulator by showing the
$I$-$V$ characteristics that are obtained by artificially setting
$\psi_i=0$ for all $i$. In this case, no electric field exists in the
central part so that the modification of the wave functions does not
occur. The $I$-$V$ curves do not show any $L_C$ dependence apart from
the fine structures coming from the discreteness of the energy
spectrum. We obtain the threshold bias voltage $V_{\rm th}\sim \Delta$
regardless of the length $L_C$ as in band insulators.

\begin{figure}
\includegraphics[height=6.6cm]{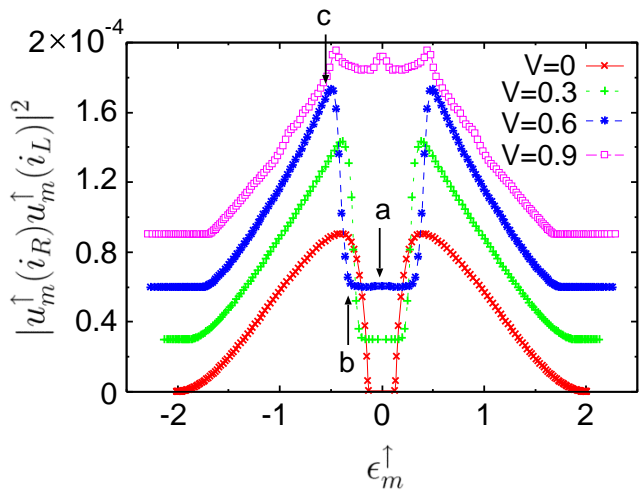}
\caption{(Color online) $|u^{\sigma}_m(i_R)u^{\sigma}_m(i_L)|^2$ with
 $\sigma =\uparrow$ plotted against $\epsilon^{\uparrow}_m$ for $V=0$,
 $0.3$, $0.6$, and $0.9$ in the case of $\delta t =0$, $L_C=200$,
 $U=1.5$, $V_p=0.1$, and $\gamma_L=\gamma_R=0.1$. For $V=0.3$, $0.6$,
 and $0.9$, the lines are shifted upward by $0.3$, $0.6$, and $0.9$,
 respectively.}
\label{fig:mott_uiRuiL}
\end{figure}

In Fig. \ref{fig:mott_uiRuiL}, we show 
$|u^{\uparrow}_m(i_R)u^{\uparrow}_m(i_L)|^2$ as a function of the real
part of the one-particle energy $\epsilon^{\uparrow}_m$ for several
values of $V$, where each line for $V\neq 0$ is shifted upward by
$V$. The behavior of this quantity for $\sigma=\downarrow$ is the same. 
\begin{figure}
\includegraphics[height=9.0cm]{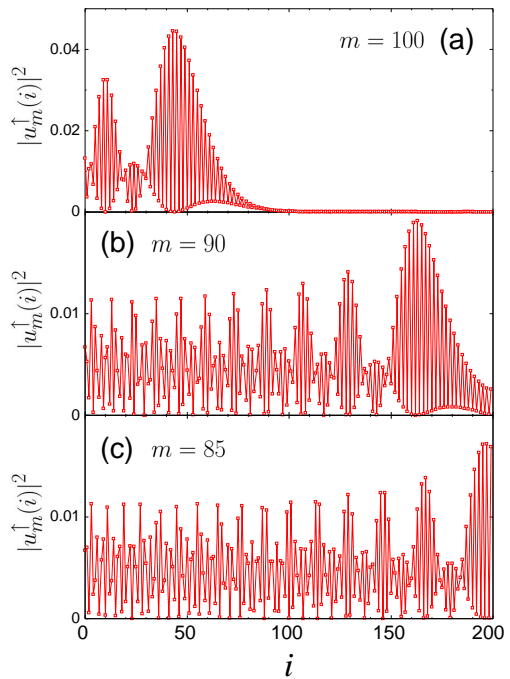}
\caption{(Color online) Spatial dependence of one-particle states
 $|u^{\uparrow}_m(i)|^2$ for (a) $m=100$, (b) $m=90$, and (c) $m=85$ in
 the case of $V=0.6$. The other parameters are the same as in
 Fig. \ref{fig:mott_uiRuiL}. The corresponding $\epsilon^{\uparrow}_m$
 are indicated by the arrows in Fig. \ref{fig:mott_uiRuiL}.}
\label{fig:mott_wf}
\end{figure}
For $V=0$, no state exists in the region 
$-\Delta/2<\epsilon^{\sigma}_m<\Delta/2$ since the energy gap opens.
$|u^{\uparrow}_m(i_R)u^{\uparrow}_m(i_L)|^2$ shows two bands that
correspond to the upper and lower Hubbard bands.
When $V=0.3$, one-particle states leak from the electrodes to the
central part so that several
states appear in the region $-\Delta/2<\epsilon^{\uparrow}_m<\Delta/2$.
As in band insulators, these states do not have any contributions to the
current because their $|u^{\uparrow}_m(i_R)u^{\uparrow}_m(i_L)|^2$ are
vanishingly small. As we increase $V$ further, e.g., $V=0.6$, the states
with vanishingly small $|u^{\uparrow}_m(i_R)u^{\uparrow}_m(i_L)|^2$
appear in a wider range around $\epsilon^{\uparrow}_m =0$. The number of
these localized states also increases. The appearance of the localized
states keeps the central part resistive until $V/L_C$ reaches the
threshold electric field $E_{\rm th}$ even if $V>\Delta$ holds. When the
system is conductive ($V>V_{\rm th}$), the upper and lower
Hubbard bands are merged into a single metallic band. In this case, all
the states around $\epsilon^{\uparrow}_m \sim 0$ are delocalized and
contribute to the current.

Figure \ref{fig:mott_wf} shows $|u^{\uparrow}_m(i)|^2$ as a function of
$i$ for $V=0.6$. Here $m$ is chosen at $100$, $90$, and $85$, whose
$\epsilon^{\uparrow}_m$ are located at the positions indicated by the
arrows in Fig. \ref{fig:mott_uiRuiL}. In Fig. \ref{fig:mott_wf}(a), we
show the wave function with 
$\epsilon^{\uparrow}_m=-0.00395$ ($m=100$), which is localized on the
left half of the central part. This state belongs to the lower Hubbard
band. Since the scalar potential is high (low) near the left (right)
electrode, 
the state is located near the top of the lower Hubbard band.
The one-particle states in the band gradually lose their localized nature
as $m$ is lowered. This can be seen in Fig. \ref{fig:mott_wf}(b) for the
case of $\epsilon^{\uparrow}_m=-0.361$ ($m=90$), where its largest
amplitude is shifted to the right compared to that of $m=100$. Figure
\ref{fig:mott_wf}(c) shows the one-particle state for 
$\epsilon^{\uparrow}_m=-0.497$ ($m=85$), which is completely
delocalized. Its wave function has large amplitudes near both
electrodes.
The spatial dependences of the one-particle states in the resistive
phase are similar to those in band insulators. 

\section{Summary}
We have investigated the $I$-$V$ characteristics of the one-dimensional
band and Mott insulators attached to electrodes. A tight binding model
with alternating transfer integrals for the band insulator and the Hubbard
model for the Mott insulator are studied by using the nonequilibrium
Green's function method. The applied bias voltage induces a breakdown
of the insulating state to convert into a conductive state for both
models. The
threshold shows a crossover as a function of the size $L_C$ of the
insulators. For $L_C\lesssim \xi =W/\Delta$, the breakdown occurs at 
$V_{\rm th}\sim \Delta$ so that the threshold is governed by the bias
voltage. For $L_C\gg \xi$, the electric field determines the threshold,
$V_{\rm th}/L_C\propto \Delta^2/W$, which is consistent with the LZ
breakdown reported previously.\cite{Oka_PRL03,Oka_PRL05}
Since the crossover is obtained for both band and Mott insulators, the
electron-electron interaction is not responsible for the
phenomenon. The profile of the scalar potential $\psi_i$, which is
linearly increasing with $i$ in the resistive phase so that the
electrons in the central part feel an almost uniform electric field, is
important for the realization of the LZ breakdown and the crossover
behavior. 

\begin{acknowledgments}
This work was supported by Grants-in-Aid for Scientific Research (C)
 (Grant No. 19540381) and Scientific Research (B) (Grant No. 20340101),
 and by ``Grand Challenges in Next-Generation Integrated Nanoscience''
 from the Ministry of Education, Culture, Sports, Science and Technology
 of Japan.
\end{acknowledgments}

\appendix*
\section{}

\begin{figure}
\includegraphics[height=5.7cm]{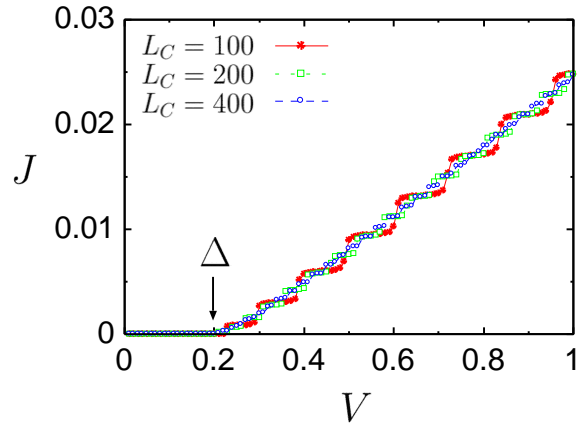}
\caption{(Color online) $I$-$V$ characteristics in the case of $\delta
 t=0.05$, $U=0$, $\gamma_L=\gamma_R=0.1$, and $\psi_i=0$ for several
 values of $L_C$. The arrow indicates the location of the gap $\Delta$.}
\label{fig:band_psi0}
\end{figure}

In this appendix, we show the results when we artificially set
$\psi_i=0$ for all $i$. The $I$-$V$ curves of the band
insulator with $\delta t=0.05$, $U=0$, $\gamma_L=\gamma_R=0.1$, and
$\psi_i=0$ for $L_C=100$, $200$, and $400$ are shown in
Fig. \ref{fig:band_psi0}. It is apparent that the breakdown occurs at 
$V_{\rm th}\sim \Delta$ regardless of the length $L_C$, 
which is consistent with Ajisaka {\it et al.}\cite{Ajisaka_PTP09} 
This is in contrast to the results in Fig. \ref{fig:band_cross} where
$V_{\rm th}$ is proportional to $L_C$ for $L_C\gg\xi$.

\begin{figure}
\includegraphics[height=5.7cm]{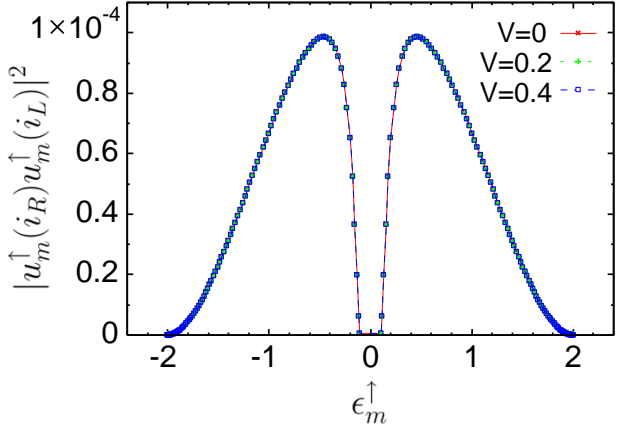}
\caption{(Color online) $|u^{\sigma}_m(i_R)u^{\sigma}_m(i_L)|^2$ with
 $\sigma =\uparrow$ plotted against $\epsilon^{\uparrow}_m$ for $V=0$,
 $0.2$, and $0.4$ in the case of $L_C=200$. 
The other parameters are the same as in Fig. \ref{fig:band_psi0}.
}
\label{fig:band_App_uiRuiL}
\end{figure}

When we fix $\psi_i=0$, one-particle states do not leak from the
 electrodes to the central part since the electric field is absent from
 the central part. In Fig. \ref{fig:band_App_uiRuiL}, we plot
 $|u^{\uparrow}_m(i_R)u^{\uparrow}_m(i_L)|^2$
as a function of $\epsilon^{\uparrow}_m$ for several values of $V$.
The results indicate that the one-particle energies and the wave
 functions are not affected by $V$. This comes from the fact that the
 Hamiltonian in Eq. (1) does not depend on $\langle n_i\rangle$ for
 band insulators if we set $\psi_i=0$ for all $i$. Therefore, no
 localized state appears inside the gap for $V=0$. In this
case, the current begins to flow when $V$ merely exceeds 
$V_{\rm th}\sim \Delta$ since one-particle states with finite
 $|u^{\uparrow}_m(i_R)u^{\uparrow}_m(i_L)|^2$ appear in the region
 $\mu_R<\epsilon^{\uparrow}_m<\mu_L$.

\begin{figure}
\includegraphics[height=5.7cm]{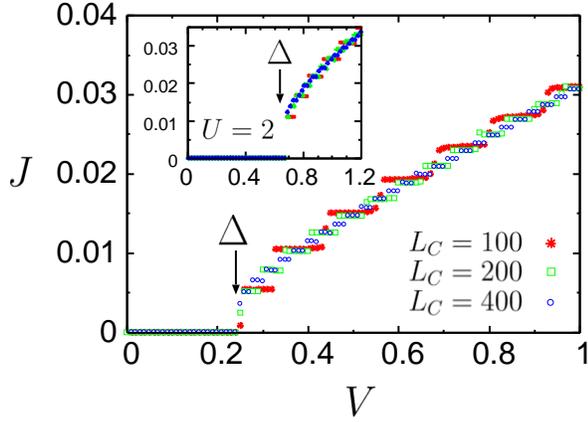}
\caption{(Color online) $I$-$V$ characteristics in the case of $\delta
 t=0$, $U=1.5$, $\gamma_L=\gamma_R=0.1$, and $\psi_i=0$ for several
 values of $L_C$. The results for $U=2$ and $\psi_i=0$ are also shown in
 the inset. The arrow indicates the location of the gap $\Delta$.}
\label{fig:mott_psi0}
\end{figure}

In Fig. \ref{fig:mott_psi0}, we show the $I$-$V$ curves of the
Mott insulator with $\delta t=0$, $U=1.5$, $\gamma_L=\gamma_R=0.1$, and
$\psi_i=0$ for $L_C=100$, $200$, and $400$. As in the case of
band-insulators, the breakdown occurs at $V_{\rm th}\sim \Delta$ for
all $L_C$. Although the breakdown seems to be continuous for $L_C=100$
and $200$, a small discontinuity is evident for $L_C=400$, which
indicates a first-order transition. The discontinuity is more obvious
for large $U$ as shown in the inset of Fig. \ref{fig:mott_psi0} for
$U=2$ with the gap $\Delta=0.68$.


\end{document}